\documentclass[a4paper,fleqn]{amsart}
\usepackage{amsaddr}
\usepackage{graphicx} 
\usepackage{amsmath} 
\usepackage{color}

\usepackage{graphicx}
\usepackage{subfig}
\usepackage{multirow}

\usepackage{array,booktabs}

\usepackage{amsmath,amsfonts,amssymb}
\newcommand{\mR}{\mathbb{R}}
\newcommand{\mE}{\mathbb{E}}

\usepackage{totcount}
	\regtotcounter{table} 	
	\regtotcounter{figure} 	

\newcommand\wordcount{
   \immediate\write18{texcount -sum -1 \jobname.tex > count.txt} \input{count.txt} }

\usepackage{mathptmx}

\def\RR{\hbox{I\kern-.2em\hbox{R}}}

\begin{document}


\title{
  \fontsize{14}{14pt} \bf 
Non-parametric estimation of mixed discrete choice models\footnote{Paper presented at the International Choice Modelling Conference (ICMC2019) in Kobe, Japan.}}
	
\author{Dietmar Bauer}
\address{Bielefeld University, Econometrics}
\email{Dietmar.Bauer@uni-bielefeld.de}

\author{Sebastian B\"uscher}
\address{Bielefeld University, Econometrics}
\email{Sebastian.Buescher@uni-bielefeld.de}

\author{Manuel Batram}
\address{Bielefeld University, Econometrics}
\email{Manuel.Batram@uni-bielefeld.de}

  \medskip\noindent
\begin{abstract} 
In this paper, different strands of literature are combined in order to obtain 
algorithms for semi-parametric estimation of discrete choice models that include the modelling 
of unobserved heterogeneity by using mixing distributions for the parameters defining the preferences. The models use the 
theory on non-parametric maximum likelihood estimation (NP-MLE) that has been 
developed for general mixing models. The expectation-maximization (EM) techniques used in the NP-MLE literature are combined with strategies for 
choosing appropriate approximating models using adaptive grid techniques. 
\\
Jointly this leads to techniques for specification and estimation that can be used to obtain a
consistent specification of the mixing distribution. Additionally, also algorithms for the estimation are developed that help to decrease problems due to the curse of dimensionality. 
\\
The proposed algorithms are demonstrated in a small scale simulation study to be useful for the specification and estimation of mixture models in the discrete choice context providing some information on the specification of the mixing distribution. 
The simulations document that some aspects of the mixing distribution such as the expectation can be estimated reliably. 
They also demonstrate, however, that typically different approximations to the mixing distribution lead to similar values of the likelihood and hence are hard to discriminate. 
Therefore it does not appear to be possible to reliably infer the most appropriate parametric form for the estimated mixing distribution. 

\end{abstract}
\maketitle

\pagestyle{plain}
  
\section{Introduction}
In order to account for unobserved heterogeneous preferences of decision makers, a number of different discrete choice models have been proposed in the literature, including mixed multinomial logit models (MMNL) \cite{Train}. 
A related class of models are multinomial probit models (MNP) with random coefficients which are often used in the panel data context (see e.g. \cite{bhat2011maximum}). 
 
When using these mixture models the modeler needs to specify the mixing distribution. 
In this respect the literature does not provide much guidance: \cite{Train} only provides a number of choices (including uncorrelated normal, log-normal and tent distributions) that are also implemented in software, both in MATLAB \cite{Train_MATLAB} as well as in R for example in the packages {\tt mlogit} or {\tt apollo} \cite{Sarrias2017,hess_apollo}. 
For MNP modelling almost only Gaussian mixture densities are used probably due to the fact that this does not complicate the estimation process (see below). 
Guidance for choosing between those models is not readily available apart from comparing the achieved likelihood values in combination with a trial and error strategy. 

A non-parametric alternative based on grid methods has been proposed for MMNL models in \cite{train2016mixed}.  
This approach is closely related to latent class models.
This has subsequently been refined as the so called logit mixed logit model (LML) in \cite{bansal2018minorization}.
A similar idea has also been applied in the MNP case by \cite{bhat2018new}. 
The usage of such non-parametric models leads to identifiability questions that are somewhat different from the parametric case. 

Contrary to the discrete choice literature there exists a large body of papers in the general statistics literature dealing with non-parametric maximum likelihood estimation (NP-MLE) for mixed models, see for example the survey \cite{Lindsay}. 
This literature is directed mostly towards situations where only a few parameters are mixed. 
It covers fully non-parametric cases as well as mixed cases where some parameters are mixed and others are treated as fixed (similar to the LML case, \cite{bansal2018minorization}). 
We discuss parts of this literature in section~\ref{sec:NP-MLE}. 
One result of this strand of literature is that the NP-MLE is given by a discrete distribution supported only on a relatively small number of grid points, which is helpful for the estimation. 
Here, small is in the sense of being roughly equal to the sample size; empirical evidence hints often to much smaller numbers.

The usage of a fixed grid is not very flexible and suffers from the curse of dimensionality, that is with increasing dimension of the parameter vector it becomes increasingly difficult to cover the whole parameter set. 
Adaptive grids will be used in this paper in order to direct the search for finding the most relevant support points. 

Thus the contribution of this paper is to combine the existing literature on the estimation of mixed discrete choice models (that is also dealt with for example in \cite{bansal2018minorization}) with an approach for adaptive grid choice 
in order to achieve estimation procedures that provide via their non-parametric nature
information on the mixing distribution also in situations of moderate dimensions of the
mixed parameters. A minor second contribution is to suggest a different type of
approximation of the underlying mixing distribution related to kernel density
estimators.

The paper is organized as follows: 
In the next section we describe the various models and discuss the respective advantages and problems. 
In section~\ref{sec:NP-MLE} we connect the models to the literature on NP-MLE.
Section~\ref{sec:algo} proposes estimation algorithms, which are adapted to situations with a moderate number of mixed parameters in section~\ref{sec:hd}. 
The estimation algorithms are demonstrated in a small scale simulation study in section~\ref{sec:simu}. 
The paper concludes with a discussion of further research needed in section~\ref{sec:concl}.

\section{Models for unobserved heterogeneity} \label{sec:models}
In many different contexts, it has been observed that parametric models 
often underestimate the variability that is present in data. 
In the context of discrete choice modelling within the random utility paradigm, this has been interpreted as the effect of heterogeneous preferences between different decision makers. 
This heterogeneity is not directly observed and typically only partly modelled using exogenous regressors. 
Instead unobserved heterogeneity is modelled as random variations in parameters. 

In more detail assume that the choice of the $i$-th decision maker in her $t$-th decision for one out of $J$ possible alternatives is modelled using a random utility model (RUM) implying a choice probability depending on underlying characteristics of the decision and the decision maker $X_{i,t} \in \mR^K$ and a parameter vector $\beta$ such that the probability, that the choice $y_{i,t}=j$ is taken, is given as:
\begin{equation} \label{eq:prob} 
{\mathbb P}(y_{i,t} = j| X_{i,t}; \beta) = p(y_{i,t} =j| X_{i,t}, \beta),
\end{equation} 
for some function $p(.|.)$ specific to the RUM formulation. 
Examples for such RUM choice models are the multinomial logit (MNL) and the multinomial probit (MNP) model and their extensions (such as generalized extreme value models etc.), which are well documented in the literature \cite{Train}. 

In this setting heterogeneous preferences can be described by letting the parameter vector $\beta$ be specific for the $i$-th individual such that ${\mathbb P}(y_{i,t} = j| X_{i,t}; \beta_i) = p(y_{i,t} =j| X_{i,t}, \beta_i)$. 
In this panel setting the parameter $\beta_i$ can only be estimated consistently if $T \to \infty$ such that every decision maker is observed for a large number of choices. 
In most applications this is unrealistic as typically $T$ is relatively small ranging up to at most a dozen choices. 

An alternative approach is to segment the population into classes with homogeneous preferences resulting in $\beta_i \in \{ b_s, s=1,...,S \}$. If the class membership would be known, we could estimate the parameters $b_s$ from repeated observations in each class. 
If the membership is unknown we obtain for the observed choices the corresponding marginal choice probabilities 
\begin{equation*}
p(y_{i,t} = j| X_{i,t}; \beta) = \sum_{s=1}^S p(y_{i,t} =j| X_{i,t}, b_s) \pi_s,
\end{equation*}
where the mixing coefficients $\pi_s > 0, ~ \sum_{s=1}^S \pi_s = 1$ denote the relative frequencies of the preferences being present in the considered population. 
Such models have been termed {\em latent class models} (see, e.g., \cite{Train}, p. 139 and the references contained therein). 

The number of classes that can be identified depends on the properties of the regressor variables $X_{i,t}$. 
Clearly, if $X_{i,t}=X$ does not vary at all, no latent classes can be identified. 
\cite{Grun2008} show that in many cases the regressor variables need to take on at least $2S-1$ different values in order to identify $S$ latent classes for a correctly specified function $p(.|.)$. 

If the regressors $X_{i,t}$ contain continuously varying parameters such as distance, price, travel time (as is typical for mode choice problems in transportation for example) then latent class models with an arbitrary -- but finite -- number of classes can be identified asymptotically. 
This holds true for MNL and MNP models as well as their extensions. 
This follows for example from the theory of identifiability of neural nets \cite{Sontag}. 

In the discrete choice literature the mixing distribution often is not modelled as a discrete-valued random variable leading to latent classes but as a continuous random variable leading to a distribution of preferences within the population such that the marginal choice probabilities are obtained by marginalizing the conditional distribution:
\begin{equation*} 
p(y_{i,t} = j| X_{i,t}; \beta) = \int p(y_{i,t} =j| X_{i,t}, b) f(b) db,
\end{equation*}
where the probability distribution function $f(b)$ is characterized for example via its mean $b_s$ and its variance matrix $\Sigma_s$ or some other parameters collected in the vector $\beta$. 
\cite{Train} lists the normal, log-normal and tent-distributions as possible forms.
These are also implemented in software, for example in the MATLAB package by Kenneth Train and the R-packages by \cite{Sarrias2017} and \cite{hess_apollo}. 

If in this model the density $f(b)$ is modelled parametrically then the parameters typically are identified if the support of the regressor space contains an open subset of $\mR^K$ and if the parametric density can be identified from its moments (in the univariate case this holds for example for the normal and the log-normal distribution as well as for the tent distribution; in the multivariate case it is simple to show for independent components and the above mentioned distributions or for the multivariate normal distribution). 
For the MNL case this follows for instance from the proof contained in \cite{fox2012random}. 
In the MNP case similar arguments can be used for normally distributed mixing coefficients.   

Furthermore \cite{Fox2017} states that if at least one component of the regressor vector is supported in $\mR$ and the corresponding coefficient has the same sign for all decision makers (as would be intuitively plausible for monetary costs in most choice models) then the joint distribution of both the random coefficient as well as the random error term in the RUM formulation can be identified. 
This allows the identification (and thus estimation) of very general models that only encode the relation between the regressors and the random utility parametrically but leave the density of the random errors totally unspecified. 

As any continuous distribution can be approximated arbitrarily close (in the sense of maximal distance of the corresponding distribution functions tending to zero) by a mixture of point masses with corresponding weights $\pi_s$ a general model for mixed RUMs can be given as
\begin{equation} \label{eq:mix_general} 
p(y_{i,t} = j| X_{i,t}; \beta) = \sum_{s=1}^S \pi_s \int p(y_{i,t} =j| X_{i,t}, \beta_s) dF_s(\beta_s),
\end{equation} 
where $F_s(\beta_s)$ denotes a cumulative distribution function relating either to a point mass or to a specific distribution such as the multivariate normal. 

The approach of \cite{Train} refined in \cite{bansal2018minorization} is to model $\pi_s$ as a logistic function of some underlying characteristics of the parameter vectors $\beta_s$. 
This reduces the problems of the curse of dimensionality but introduces the problem of specifying the variables determining the mixing as well as the functional form for the mixing. 

Model~\eqref{eq:mix_general} also contains the approach of \cite{bhat2018new} as a special case with $p(.|.)$ denoting the MNP probability function.

\subsection{The special case of normally distributed random parameters}
The general model discussed in the last section has one major drawback: 
The integrals in \eqref{eq:mix_general} in most cases do not have a closed form solution except in special cases. 
One special case is constituted by point masses implying that the integral amounts to evaluation at the points with nonzero probability. 

A second special case is constituted by multivariate normal distributions in the context of MNP modeling for cross sections such that $t=T=1$: 
Let the random utility of alternative $j$ in the $t$-th choice of decision maker $i$ be given as 
\begin{equation*} 
U_{j,i,t} = \alpha_j + X_{j,i,t}'\beta + e_{j,i,t} = \alpha_j + X_{j,i,t}'b + (X_{j,i,t}'\tilde \beta + e_{j,i,t}),
\end{equation*} 
where the vector $[e_{j,i,t}]_{j=1,...,J}$ and the vector $\tilde \beta$ are independently multivariate normally distributed, while $b$ is deterministic. 
It follows that given $X_{j,i,t}, j=1,...,J$ the random utility is distributed multivariate normally. 
Therefore, in this case evaluation of the choice probability requires the evaluation of a $J-1$ dimensional Gaussian cumulative distribution function. 
This is the same problem as for nonrandom coefficients. 
Calculation of the Gaussian cumulative distribution function  can be done either 
by numerically approximating the integral or by using simulation based methods. 
For large $J$ both methods are time consuming. This complexity is not related to 
the mixing, as follows from above. 
Therefore, in this case the evaluation of the choice probabilities for the case of unobserved heterogeneity is no more complicated than without. 

In the MNP case it is also possible to combine different types of mixing distributions such that some parameters can be modelled to mix subject to a normal distribution while others might show other mixing properties as has been noted by \cite{bhat2018new}.

For the mixed MNL models evaluating the choice probabilities with Gaussian mixing distributions requires simulations or numerical integration methods. 
For general mixing distributions, however, this is the case for both models, MNL and MNP. 

\subsection{The panel data case} 
In the panel situation the choices are observed for a number of choice situations for the same decision maker. 
Therefore, even if the preferences vary between decision makers, they should remain constant for all decisions of one decision  maker. 
This implies correlations between the random terms across decisions. 

The correlations can be included in the model easily when coefficients $\beta$ are drawn randomly in order to approximate the integral by a Monte Carlo estimate. 
Typical software packages include such random draws leading to maximum simulated likelihood methods. 
The corresponding asymptotic theory is provided in \cite{Train}. 

Alternatively in the MNP framework the {\em MaCML} (maximum approximate composite marginal likelihood) approach has been introduced by \cite{Bhat}. 
In this approach the likelihood is replaced by the composite marginal likelihood (CML). 
The most prominent approach uses the so called pairwise likelihood which includes the probabilities of pairs of choices in the criterion function. 

In the case of the MNP model with random parameters which are multivariate normally distributed we obtain with the same argumentation as above the fact that including random parameters is no more computationally complex than evaluating the CML with only fixed parameters. 

The same holds, if normally distributed random variables are replaced by a Gaussian mixture. 
Hereby the choice probabilities are calculated as 
\begin{equation} \label{eq:mix_MNP_kernel} 
{\mathbb P}(y_{i,t_1} = j_1 \wedge y_{i,t_2} = j_2 | X_{i,t_1}, X_{i,t_2}; \beta) = \sum_{s=1}^S \pi_s \int p( y_{i,t_1} = j_1 \wedge y_{i,t_2} = j_2 | X_{i,t_1}, X_{i,t_2}; \beta_s) dF_s(\beta_s;\mu_s,\Sigma_s).
\end{equation} 
Again each of the terms for $s=1,...,S$ in the sum can be evaluated with the same code that is used for MNP models. 
Note that here we see connections to kernel density estimation where also density estimation is performed using Gaussian mixtures with fixed bandwidth, that is, with fixed variance $\Sigma = \Sigma_s$.

Again the identifiability of these structures hinges on properties of the regressor variables $X_{i,t}$, where the support of these vectors must contain an open set as a minimum requirement. Identifiability here trades off properties of the regressor vectors
with assumptions on the mixing distribution: The more flexible the mixing can be, the more restrictive the assumptions on the regressor may be, see the last section for details.


\section{Non-parametric maximum likelihood estimation} \label{sec:NP-MLE}
The non-parametric maximum likelihood estimator (NP-MLE) has been proposed by \cite{Kiefer_Wolfowitz}. 
The term NP-MLE relates to the optimization of the likelihood for the mixture over the set of all mixing distributions, that is models of the form
\begin{equation*} 
p(y_{i,t} = j| X_{i,t}; Q) = \int p(y_{i,t} =j| X_{i,t}, b) dQ(b)
\end{equation*} 
for all distributions $Q$ of the real vector $b$.  

\cite{Lindsay} provides one of the first surveys of the NP-MLE properties. 
The model structure here is very generally applicable to models with some random parameters mixed using some continuous density. 
The main property used in the literature is that the set $\Gamma = \{ [p(y_{i,t} = j| X_{i,t}; \beta)]_{i,t}, \beta \in \Theta \}  \subset [0,1]^n$ of all possible vectors $p(y_{i,t} = j| X_{i,t}; \beta)$ for parameter vectors $\beta$ ranging in some set $\Theta$ or its closure respectively is compact. 
For the mixed discrete choice models this is evident as a closed and bounded subset of a finite dimensional real vector space is compact.

As the set of latent class models is a convex set we can compute the directional derivative of the scaled log-likelihood at one mixing distribution in the direction of a point mass distribution $\delta_{\beta}$ at $\beta$ 
\begin{equation} \label{eq:D}
D(\beta;Q) = \lim_{\alpha \to 0} \left( \frac{ll_n((1-\alpha)Q +\alpha \delta_{\beta})- ll_n(Q)}{\alpha} \right) = (nT)^{-1} \sum_{i=1}^n \sum_{t=1}^T 
\left(\frac{p(y_{i,t} = j| X_{i,t}; \beta)}{p(y_{i,t} = j| X_{i,t}; Q)} -1 \right) 
\end{equation} 
where 
\begin{equation*} 
ll_n(Q) = (nT)^{-1} \sum_{i=1}^n \sum_{t=1}^T \log p(y_{i,t}| X_{i,t}; Q)
\end{equation*}
denotes the scaled log-likelihood for mixing distribution $Q$. 

The main results of interest for us relate to (a) the properties of the maximizers of the NP-MLE problem and (b) to conditions derived from the optimization.
The main messages can be summarized as follows:

\begin{itemize}
    \item There exist optimizers to the NP-MLE problem with a maximum of $nT+1$ support points. That is for some optimizer one has $\hat Q(\beta)= \sum_{s=1}^S \hat \pi_s \delta_{\beta_s}(\beta)$ with $S \le nT+1$ and hence the optimum can be obtained from a latent class model. 
    \item The optimizers fulfill the following first order conditions: $D(\beta;\hat Q) \le 0$ for all $\beta$.
    \item $D(\beta_s;\hat Q) = 0$ for each support point $\beta_s$ of $\hat Q$.
\end{itemize}

The first result shows that it is no restriction of generality to restrict the search for the mixing distribution to latent class models. 
The number of support points given is too large to be of practical use. 
However, empirically it has been verified in a number of contexts that one typically obtains good results already with a smaller number of support points.

The other two results characterize optima and provide means to identify potential new support points: Points at which $D(\beta_s,\hat Q)$ are large show potential to increase the likelihood. 
This is used in the algorithms presented in the next section.


\section{Algorithms} \label{sec:algo}
The characterization of the optimal mixing distribution obtained from the maximization of the likelihood can be guided by the evaluations of the last section. 
Consider the representation of the model in the form 
\begin{equation*}
p(y_{i,t} = j| X_{i,t}; \beta) = \sum_{s=1}^S p(y_{i,t} =j| X_{i,t}, b_s) \pi_s.
\end{equation*}
Then the estimation has to deal with three topics:

\begin{itemize}
    \item For a given number of support points $S$, select the best locations $b_s$.
    \item For given locations $b_s$ of the support points, calculate the maximizing weights $\pi_s$.
    \item Selecting an appropriate number $S$ of support points. 
\end{itemize}

These topics will be discussed in the following subsections. 
The last subsection of this section then joins these approaches to a proposed algorithm.

\subsection{Choosing locations}
The mixing distribution is characterized by the location of the support points $b_s$ as well as their corresponding weights $\pi_s$. 
For the selection of the support points two different approaches are popular: 
Either a fixed grid of points is used or the locations are estimated based on the data. 

In low (say up to three) dimensional situations a fixed grid can be used. 
In this case using $x$ points per dimension leads to a total number of $x^d$ points in the grid for $d$ dimensions. 
Thus for $d=3$ using $x=20$ points leads to a total of $8000$ support points which is manageable. 
In higher dimensions this is not feasible any more. 

The second approach uses less points but is more careful where to put them. 
A number of different methods have been used, see for example \cite{bohning1995review} for a survey. 
For the models used in this paper the classical expectation-maximization (EM) algorithm is a good option. 

The EM algorithm has been suggested for latent class models, that is a situation where optimization would be easy, if class membership of each observation would be known. 
Thus, assume that $z_{i,s}, i=1,...,n, s=1,...,S$ denotes the indicator variable for individual $i$ belonging to class $s$, in the cross sectional situation where $t=T=1$ observation per individual occur (consequently below the index for the choice occasion $t$ is dropped for notational simplicity). 
Then the full scaled log-likelihood $\tilde ll_n^F$ based on the full data set $(y_i,X_i,z_{i,s})$ differs from the likelihood for the data $(y_i,X_i)$: 
\begin{align*} 
ll_n(\beta; y_i,X_i) & = n^{-1} \sum_{i=1}^n \log ( \sum_{s=1}^S \pi_s p(y_i | X_i;b_s) ), \\
\tilde ll_n^F(\beta; y_i,X_i,z_{i,s}) & = n^{-1} \sum_{i=1}^n \sum_{s=1}^S z_{i,s} \log ( p(y_i | X_i;b_s) ).
\end{align*} 
The main change here lies in interchange of the sum and the logarithm in the full data likelihood. 
The EM algorithm uses these interchanges and proceeds in two steps (cf. also \cite{train2008algorithms}): 
\begin{itemize}
    \item[E-step:] Calculate the conditional expectation $\mE (z_{i,s} | y_i, X_i)$. 
    This expectation amounts to the calculation of conditional probabilities as $z_{i,s}$ are indicator variables. 
    We obtain the estimate $\hat \gamma_{i,s}$ of $\mE (z_{i,s} | y_i, X_i)$ as
    \begin{equation*}
    \hat \gamma_{i,s} = \frac{ \pi_s p(y_i | X_i;b_s)}{\sum_{s=1}^S \pi_s p(y_i | X_i;b_s)}. 
    \end{equation*}
    \item[M-step:] Given $\hat \gamma_{i,s}$ maximize the function
    \begin{equation*}
    \mE [\tilde ll_n^F(\beta; y_i,X_i,z_{i,s}) | y_i,X_i]  =   n^{-1} \sum_{i=1}^n \sum_{s=1}^S \hat \gamma_{i,s} \log ( p(y_i | X_i;b_s) )
    \end{equation*}
    with respect to $b_s$. 
    If all parameters are mixed, this optimization can be performed for each vector $b_s, s=1,...,S$ independently using a weighted likelihood function using standard software. If both fixed and varying parameters are present, more elaborate schemes along the lines of \cite{bansal2018minorization} can be used. 
    The weights $\pi_s$ are adapted either by using the updating formula of \cite{train2008algorithms} or using the algorithm in the next section.     
\end{itemize}

The two steps guarantee that the likelihood value $ll_n$ is increased after the M-step is completed. 
The EM algorithm then consists in iterating these two steps. \\
The EM algorithm is known to produce consistent estimators for convex likelihoods and else converges to stationary points. 
It is also known to be converging slowly compared to gradient type methods. 
However, in particular for latent class models it is easy to implement.

\subsection{Selecting weights for given locations} 
For given support points the task of estimating $\pi_s$ can be done by solving the following problem:
\begin{equation*}
\max_{\pi_s \ge 0, \sum_{s=1}^S\pi_s = 1} ll_n(Q;y_i,X_i) \quad \mbox{where} \quad 
ll_n(Q;y_i,X_i) = n^{-1}\sum_{i=1}^n \log ( \sum_{s=1}^S \pi_s p(y_i| X_i;b_s)),
\end{equation*}
which mathematically is a constrained optimization problem with linear equality constraints as well as non-negativity constraints. 
For such optimization problems very efficient estimation algorithms are implemented e.g. in the MATLAB optimization toolbox. 
These algorithms find the optimal weights sequence $\pi_s$ over the simplex fast (in particular as the gradient can be calculated analytically). 

\cite{Koenker2014} proposes to solve the dual problem rather than the primal as given above. 
However, our experience is that in typical sample sizes for discrete choice analysis with a reasonable number of support points (ranging up to several thousands) calculations in the primal problems are feasible. 

While the EM algorithm also results in new choices $\hat \pi_s = n^{-1}\sum_{i=1}^n \hat \gamma_{i,s}$ for the weights, solving the optimization problem improves in our experience in many cases the fit significantly while not contributing much to the total run time.

\subsection{Adding and removing support points} 
Up to now the number of support points has been assumed to be fixed and given. 
In practice this is not the case and methods for adjusting the number of support points have to be found. 

New support points can be found in many ways, popular strategies involve:
\begin{itemize}
    \item Random sampling: new points are drawn from some underlying probability distribution, often a multivariate Gaussian distribution with some fixed variance.
    Such draws appear only to be relevant in early stages without any prior knowledge. 
    \item Weight based re-sampling: support points showing a high weight $\pi_s$ potentially are located in areas where the mixing distribution is large. 
    Hence adding points in such regions might increase the 'resolution' of the estimated distribution.
    \item Criterion based re-sampling: new support points might be obtained from some criterion indicating promising locations. 
    In the current context a promising candidate is given by the gradient $D(\theta;\hat Q)$ defined in~\eqref{eq:D}. 
    At the optimum one has $D(\beta;\hat Q)=0$ whereas a positive value indicates that the likelihood can be increased in this direction. 
    Therefore Metropolis-Hastings sampling using this function as the criterion function can be used. 
    This sampling algorithm hence selects a new point with a large probability in regions where the likelihood can be improved a lot while regions with negative values of $D$ are not visited.  
\end{itemize}

After adding a new support point it is necessary to estimate a new value $\hat \pi_{S+1}$ and to adjust all others. 
This can be done using line-search methods for the new mixing distribution $\alpha \delta_{\beta_{S+1}} + (1-\alpha)\hat Q$. 
Line search here can be executed numerically extremely fast as the main required data $p(y_{i}|X_i;\beta_{S+1})$ and $p(y_i|X_i; \hat Q)$ are already calculated for the evaluation of $D(\beta_{S+1};\hat Q)$. 

Removing points may be done by simply dropping all points with corresponding weights $\hat \pi_s$ estimated too small, for example smaller than a small threshold like $\epsilon_{tol}=0.001$.

\subsection{Proposed Algorithm} 
\cite{Wang2013} propose a general algorithm for estimating mixture models which combines the elements from the last subsections to the following numerical scheme based on an initial estimate using $m$ support points. 
The algorithm repeats the following five steps until convergence: 

\begin{itemize}
    \item[Step 1] Run $n_{EM}=5$ steps of the EM-algorithm.
    \item[Step 2] Draw $n_g$ new support points using the Metropolis-Hastings algorithm for the function $D(\beta;\hat Q)$.
    \item[Step 3] Group the new support points into $m$ groups $C_j$ based on distance to the support points with respect to a randomly chosen component of the vectors. 
    \item[Step 4] Iteratively choose from each of the $m$ groups the one corresponding to the largest value of $D(\beta;\hat Q)$ and estimate the corresponding $\alpha$ by line search. 
    \item[Step 5] Re-estimate the weight sequence as $\hat \pi_s$ and drop points such that $\hat \pi_s \le \epsilon_{tol}$.
\end{itemize}

The algorithm contains a number of choices: 
The number of EM-steps taken in Step 1 can be changed at will. 
\cite{Wang2013} state that in their experiments 5 iterations worked well. 
Note, however, that in our setting the EM-algorithm is more costly than in other settings as the maximization cannot be done analytically and hence numerical optimization needs to be employed. 
In the application of \cite{Wang2013} for example the E and the M-step can be stated explicitly involving almost no computational cost. 

Secondly, the number $n_g$ of points chosen in each step can be tuned. 
In our implementation we use $n_g=100$. 
Note, however, that in each step the number of support points is at most doubled. 

Finally the tolerance level controls the number of support points: 
At most $1/\epsilon_{tol}$ points may fulfill this restriction. 
In practice the number is much smaller. 
Thus the addition and pruning effectively keeps the number of support points lower than an adjustable (by choosing $n_g$ and $\epsilon_{tol}$) limit.

\section{Adaptations for high-dimensional parameter sets} \label{sec:hd} 
The algorithm described in the last section has been shown to be successful in the context of estimating the distribution using Gaussian mixtures as an alternative to kernel density estimation. 
The demonstration examples of \cite{Wang2013} have dimension between 2 to 13. 

However, the situation there is simpler than our setting, as for density estimation the range of the support points is determined by the location of the data. 
In our case the mixed parameters are linked to the location of the data only indirectly via the function $p$ .

For density estimation in multivariate settings it is common to use the same bandwidth parameter in each parameter component. 
In our setting, even after standardization of the regressor variables, it is not clear whether this is a good strategy. 
Also the two extremes of latent class models (which would require a small bandwidth to approximate the point masses) and one single Gaussian distribution (requiring one large bandwidth parameter) should both be accommodated. 

\cite{Wang2013} propose to use information type criteria to select the best bandwidth parameters. 
In our setting the 'bandwidth' choice occurs by approximating the mixing distribution using a mixture of Gaussians as 
\begin{equation*}
Q(\beta) = \sum_{s=1}^S \pi_s \Phi(\beta; \beta_s,\Sigma_s).
\end{equation*}
While the $\beta_s$ determine the location of the mass centers, the matrices $\Sigma_s$ determine the concentration of the $s$-th component. 
As is usual we will use diagonal matrices $\Sigma_s = \mbox{diag}(\sigma_1,...,\sigma_K) \in \mR^{K \times K}$, but allow for different diagonal elements. Point masses are approximated using small $\sigma_j$. 

\begin{figure}
    \centering
    \begin{tabular}{cc} 
    \includegraphics[width=7.5cm]{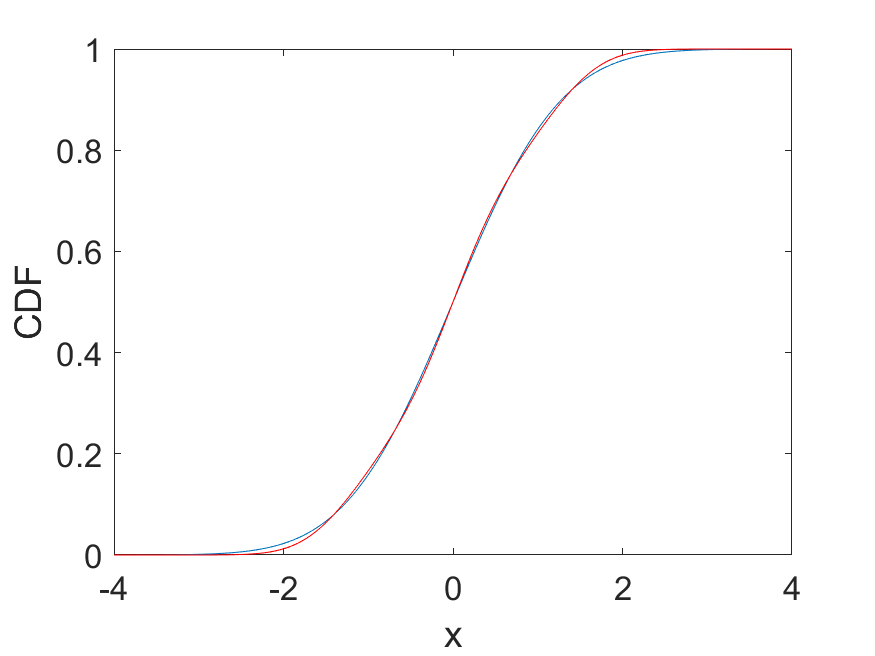} &   \includegraphics[width=7.5cm]{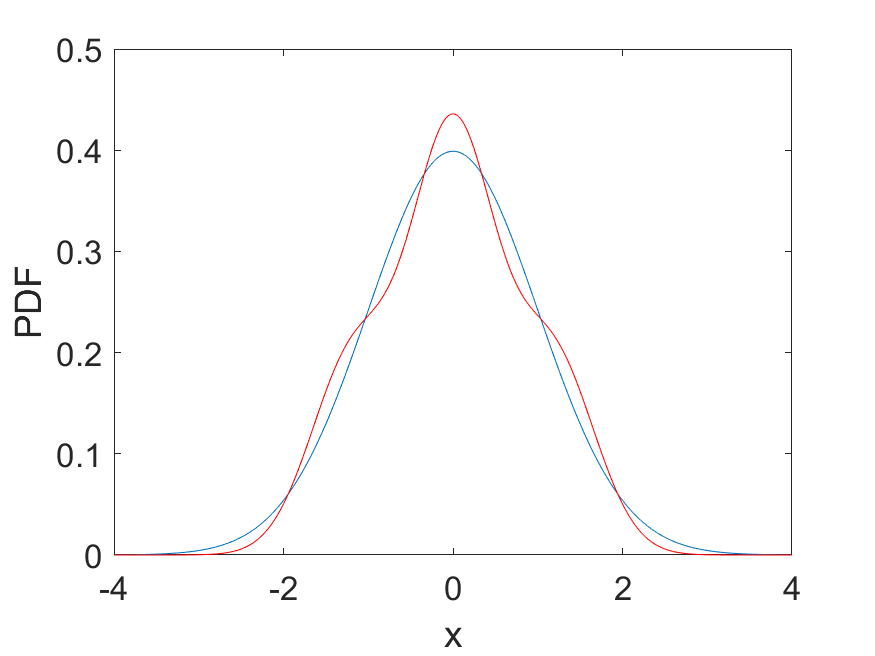} \\
    (a) CDF & (b) PDF
    \end{tabular} 
    \caption{Comparison in approximation quality: ${\mathcal N}(0,1)$ versus $0.24 {\mathcal N}(-1.18,0.5)+0.52 {\mathcal N}(0,0.5)  + 0.24 {\mathcal N}(1.18,0.5)$. }
    \label{fig:approx_cdf}
\end{figure}

This leads to a situation where in each dimension a combination of grid points characterized by location/variance pairs is obtained. 
Moreover a hierarchical structure is obtained in the sense that points with higher variance can be approximated by a number of points with smaller variance.
Figure~\ref{fig:approx_cdf} demonstrates approximating a standard normal distribution by the sum of three Gaussian random variables with variance 0.5. 
The maximal difference between the two corresponding CDFs equals 0.01, the maximal absolute difference of the pdfs equals 0.037. 

The start of the algorithm of the last section requires the selection of an initial set of support points as well as the corresponding variances $\Sigma_s$. 
This is accomplished by the following algorithm:

\begin{enumerate} 
    \item Drawing a uniform grid with a fixed relatively large bandwidth. 
    The number of grid points is adapted here such that $m = x^d$ is of order 1000, where $x$ denotes the number of grid points in one coordinate. 
    \item Then the component weights $\pi_s$ are estimated. 
    \item For the largest $m_l$ weights $\pi_s$ the corresponding components are split in all coordinates: 
    Using the above framework the uni-variate normal distribution is exchanged by three points with halved variance. 
    \item For each new potential point the gradient $D(b;\hat Q)$ is calculated. 
    If it is positive, the new point is kept, else discarded.
    \item With all points the weights are re-estimated and components with estimated weights smaller than the threshold $\epsilon_{tol}$ are discarded. 
    \item The algorithm repeats in step 3 until a stopping criterion is met. 
\end{enumerate} 

In our simulations the adaptive grid is used as a starting point and hence the simple stopping criterion of running fifteen\footnote{This setting is arbitrary. Results with five iterations 
showed almost the same results.} iterations of the adaptation is used.

\section{Simulation Study} \label{sec:simu} 
In this section a small scale simulation study is performed in order to verify that the methods proposed in this paper can be used. 
The setting is rather simple and hence the results with respect to the ranking of the various methods are only indicative.

We investigate the choice between three alternatives which are characterized by only one regressor which varies over alternatives. 
The values of the regressor specific to each alternative are drawn from independent identically distributed normal variables with expectation zero and variance 9 in order to provide a strong signal. 
The random utility is then given as
\begin{equation*}
U_{i,j} = X_{i,j}\beta + \alpha_j + \epsilon_{i,j},
\end{equation*}
where the parameters $\beta \in \mR$ and $\alpha_j, j=2,3$ may be random variables. 
We introduce the normalization $\alpha_1=0$ to fix the intercept of the utility. 
The noise term $\epsilon_{i,:} \in \mR^3$ is assumed to be distributed as multivariate normal with expectation zero and variance $\Sigma_0 = I_3$ resulting in the estimation of MNP models. 
Fixing the variance implies a standardization of the scale of the utility. The choice probabilities are evaluated using the function {\tt bvnu} in {\tt MATLAB}. 

We consider two main cases with variations in between: 
In cases 1 no ASCs are included ($\alpha_j=0$) while the coefficient $\beta$ is assumed to be random according to the following sub-cases:

\begin{itemize}
    \item[(1a)] $\beta = 1$ with probability $p=0.75$ and $\beta = -1$ otherwise. 
    \item[(1b)] $\beta$ is normally distributed with expectation 1 and variance 1.
    \item[(1c)] $\beta$ is distributed log-normally with $\mu=0$ and $\sigma = 0.5$.
\end{itemize}

For the second case the coefficient $\beta=1$ is fixed and the ASCs are estimated non-parametrically. This is equivalent to assuming that the distribution of the error terms
is changed. 
We distinguish the case (2a) of normally distributed error terms with correlations as 
\begin{equation*}
\Sigma_0  = \left[ \begin{array}{ccc} 1.00 & 0.5 & 0 \\ 0.5 & 1.25 & 0.5 \\ 0 & 0.5 & 1.25 \end{array} \right]. 
\end{equation*}
As a second case (2b) we draw the errors $\epsilon_{i,:}$ from a mixture of two normal distributions with probability 0.5 each. 
The first equals the one above, the second has expectation $(0,1,-1)'$ and variance
\begin{equation*}
\tilde \Sigma  = \left[ \begin{array}{ccc} 1.00 & -0.50 & 0 \\ -0.50 & 1.25 & -0.50 \\ 0 & -0.50 & 1.25 \end{array} \right]. 
\end{equation*}
%

In all cases we deal with four estimates:

\begin{itemize}
    \item[GR:] The adaptive grid approach described in section~\ref{sec:hd}. Here we use $\Sigma=0.1I_3$ (the $3 \times 3$ identity matrix) in the MNP specification. 
    \item[EM-GR:] The EM based algorithm of section~\ref{sec:algo} starting from the estimate {\tt GR}. Here $\Sigma=\Sigma_0$ is used.
    Five iterations of five EM steps followed by a scanning for new points are used. 
    The EM steps only adapt the location of the grid points but not the variances. 
    \item[EM:] This is identical to {\tt EM-GR} but starts from the same initial grid provided at the start of {\tt GR}. Thus in this algorithm the variance of the components remains constant. Again $\Sigma=\Sigma_0$
    \item[BE:] This chooses the best estimate (that is the one with the highest log-likelihood value) of the other three. 
\end{itemize}

The four estimates are calculated for four sample sizes $I = 500, 1000, 2500, 5000$.
For each sample size and each case $M=500$ replications are used. 

Note that this comparison is not entirely fair as {\tt GR} uses less information in that
a small variance is used in the MNP kernel, while the other methods use the true
variance for the MNP kernel. For cases (1a)-(1c) this poses a difficulty as in this case {\tt GR} does contain the true data generating process but uses a wrong scaling due to the
different error variance. 
In real
applications this is a realistic assumption, as there a compromise must be found via
estimating $\Sigma$. For cases (2a)-(2b) this issue is not that important as there ASCs are
modelled as being mixed, which can be seen as one way to model a general error variance $\Sigma$. 

The results are compared with respect to a number of different performance measures: 

\begin{itemize}
    \item The value of the scaled log-likelihood $ll_n(\hat Q;y_i,X_i)$ compared to the scaled log-likelihood $ll_{n}(Q_0;y_i,X_i)$ at the true parameters: 
    $ll_n(\hat Q;y_i,X_i) \ge ll_{n}(Q_0;y_i,X_i)$ should hold in all cases. 
    However, finding the optimal log-likelihood is not simple in all cases. 
    Thus, values of $ll_n(\hat Q;y_i,X_i) - ll_{n}(Q_0;y_i,X_i)$ smaller than zero indicate problems in the optimization. 
    \item The mean absolute distance $d_P$ between the estimated choice probabilities and the actual choice probabilities for the chosen alternatives (cross sectional case, thus $t=1$):
    \begin{equation*}
    d_P^{(E)} = \frac{1}{IM}\sum_{m=1}^M \sum_{i=1}^I | p(y_{i,1} | X_{i,1}, \hat Q^{(E)}) - {\mathbb P}_0(y_{i,1} | X_{i,1}) |,
    \end{equation*}
    where $E$ stands for the four estimators and ${\mathbb P}_0(y_{i,1} | X_{i,1})$ denotes the true conditional choice probability of the chosen alternative for individual $i$. 
    \item Kolmogorov-Smirnoff (KS) distance between the estimated mixing distribution and the actual mixing distribution: 
    For two CDFs $C_1(z)$ and $C_2(z)$ the one norm is defined as
    \begin{equation*}
    d_{CDF,1}(C_1,C_2) = \int_{z} | C_1(z)-C_2(z) | 
    \end{equation*}
    Numerically the integral is approximated as the sum over a grid of bin size 0.01 ranging from $[-4,4]$ (in all directions in the multivariate case).
    \item The percent of negative estimates of the parameter $\beta$ is compared to the true probability. In order to measure estimation accuracy we calculate the mean absolute distance of the percent negative coefficients.
\end{itemize}

\subsection{Results for Cases (1a)-(1c)} 
\begin{figure}
\begin{tabular}{c m{3.5cm} m{3.5cm}  m{3.5cm} m{3.5cm}}
Case (1a) &  \includegraphics[width=35mm]{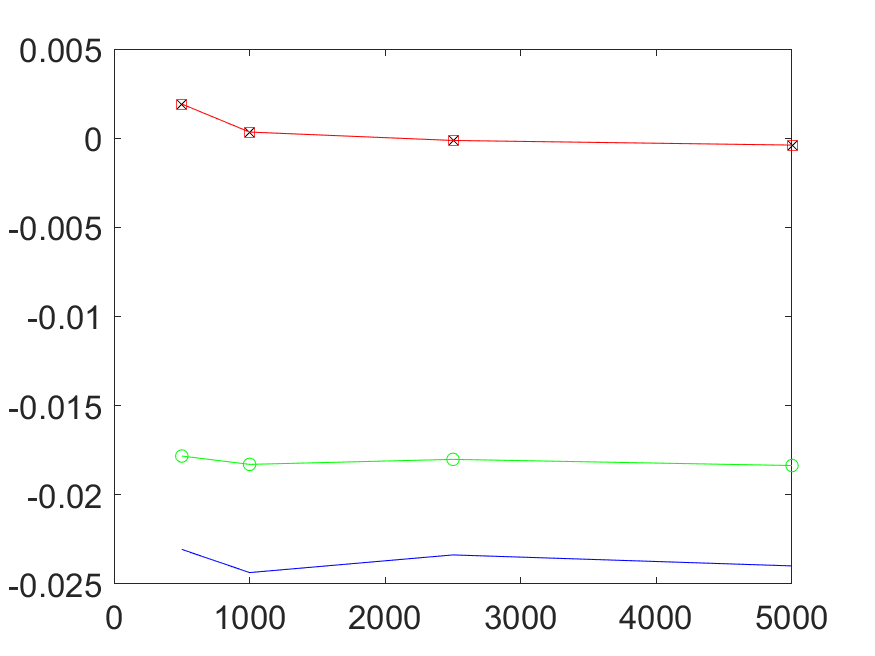} &   \includegraphics[width=35mm]{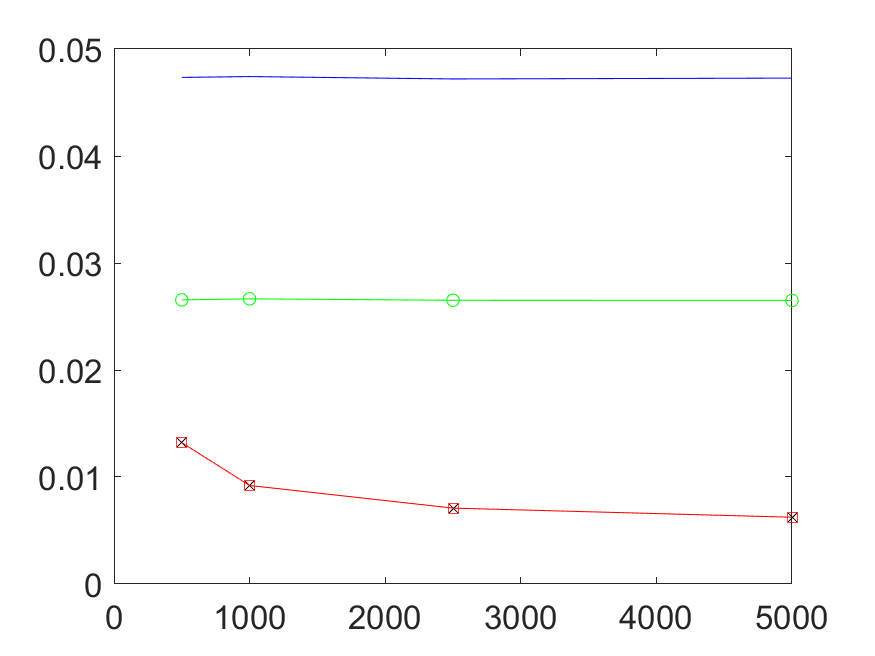} &  \includegraphics[width=35mm]{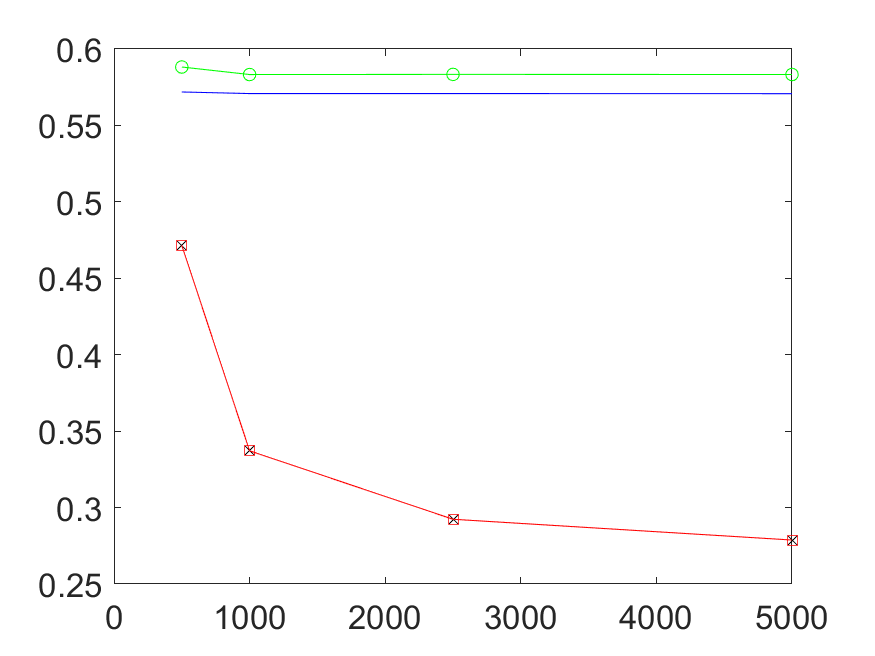} & 
\includegraphics[width=35mm]{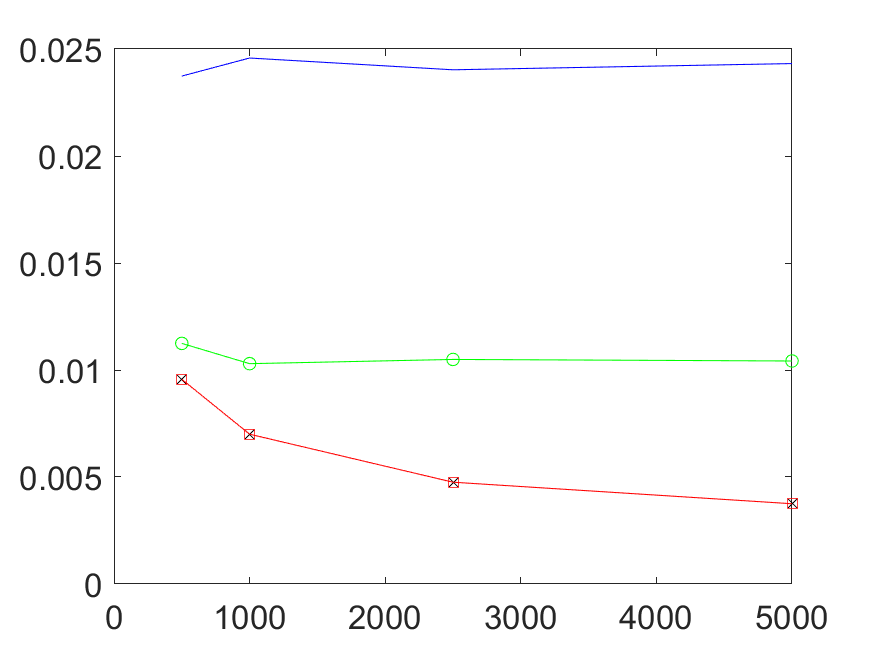}  \\
Case (1b) &  \includegraphics[width=35mm]{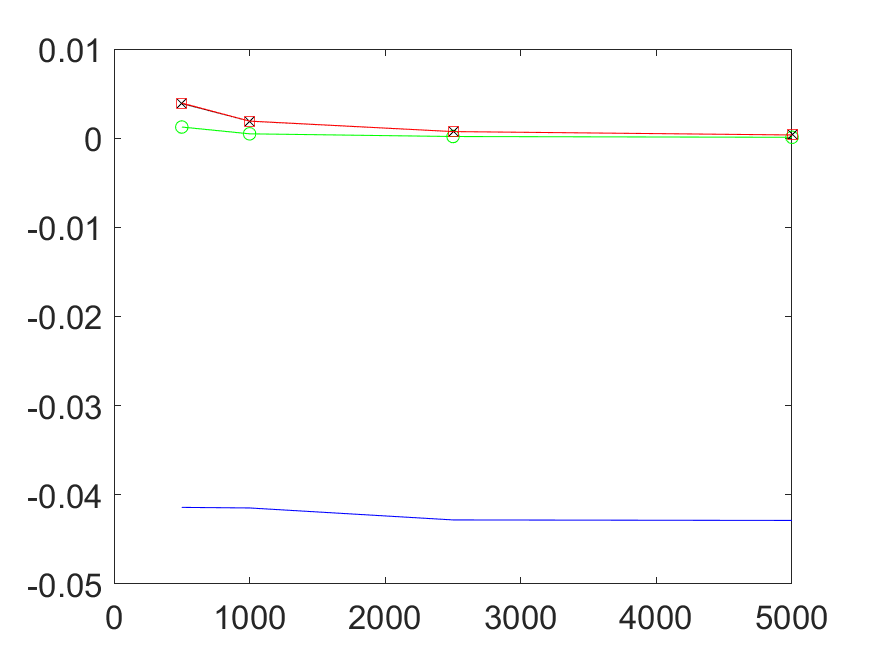} &   \includegraphics[width=35mm]{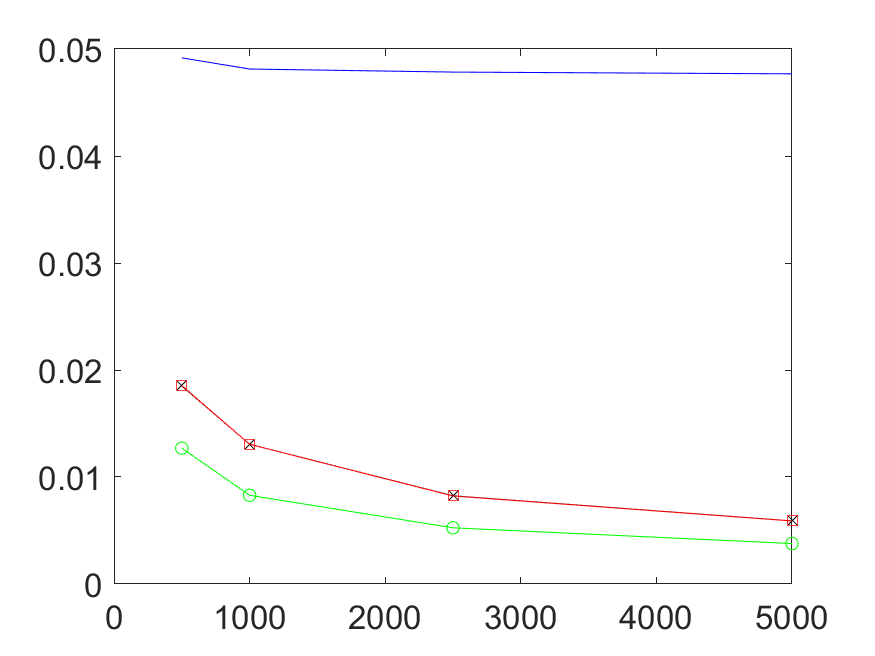} &  \includegraphics[width=35mm]{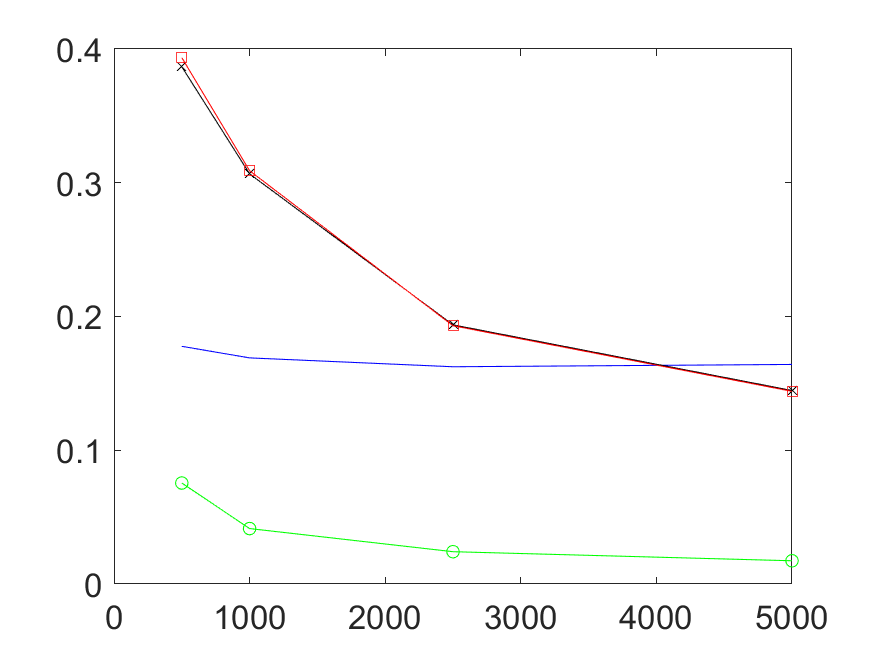}  &
\includegraphics[width=35mm]{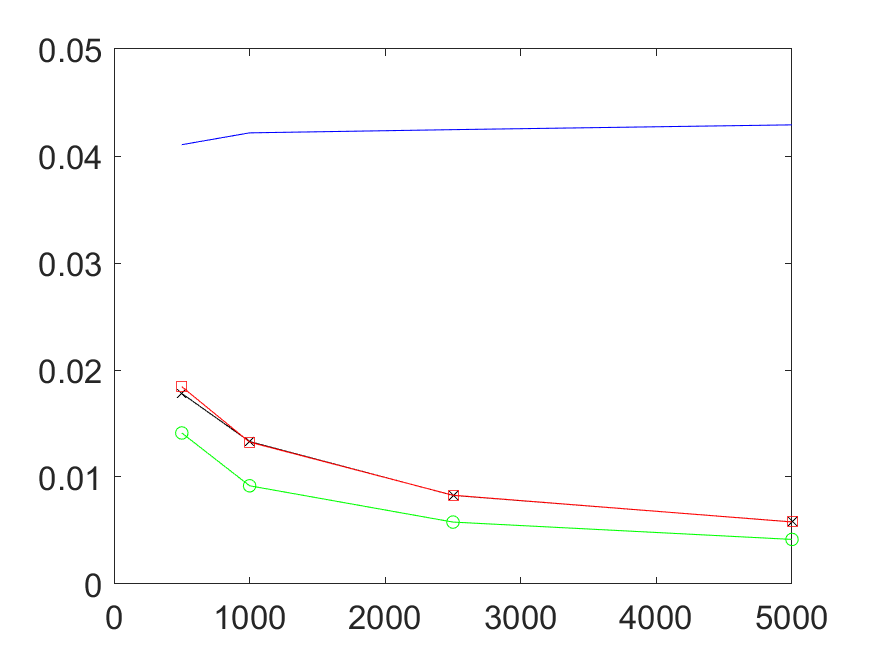} \\
Case (1c) &  \includegraphics[width=35mm]{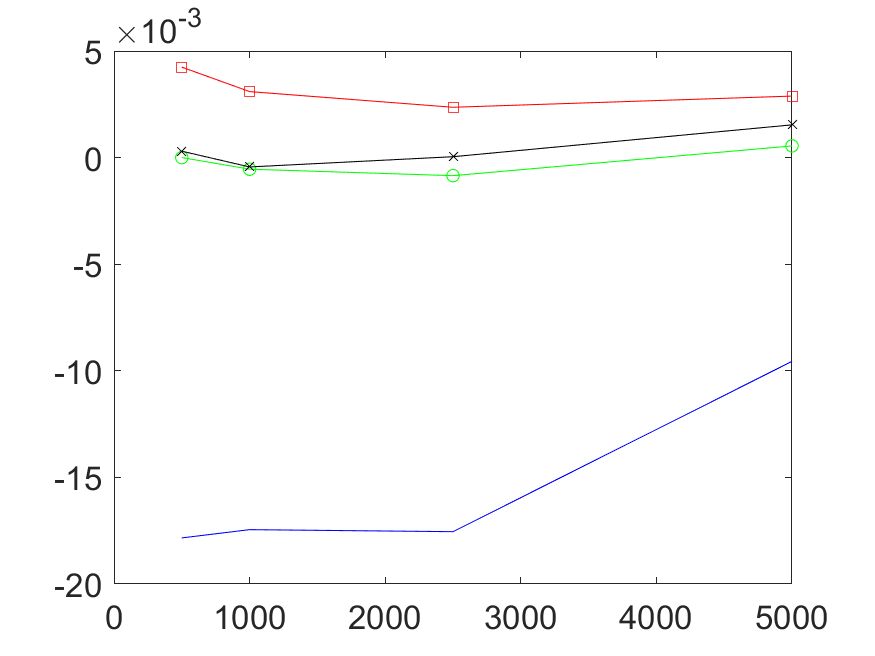} &   \includegraphics[width=35mm]{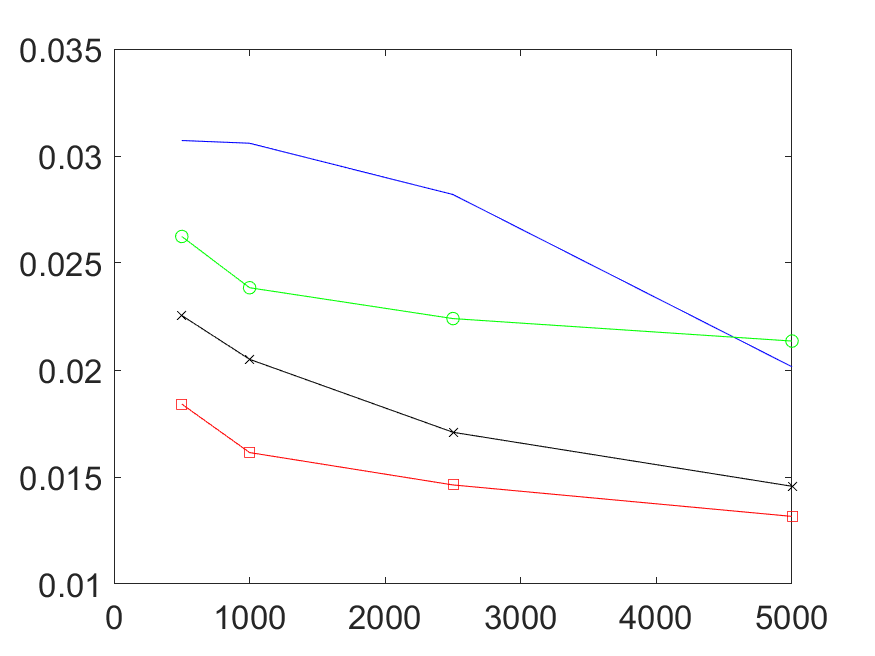} &  \includegraphics[width=35mm]{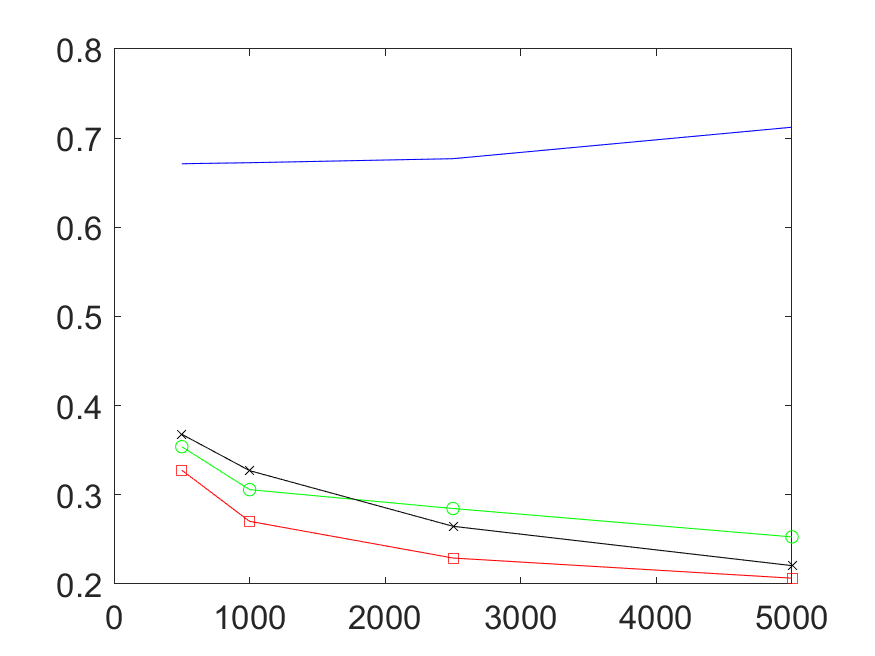} & 
\includegraphics[width=35mm]{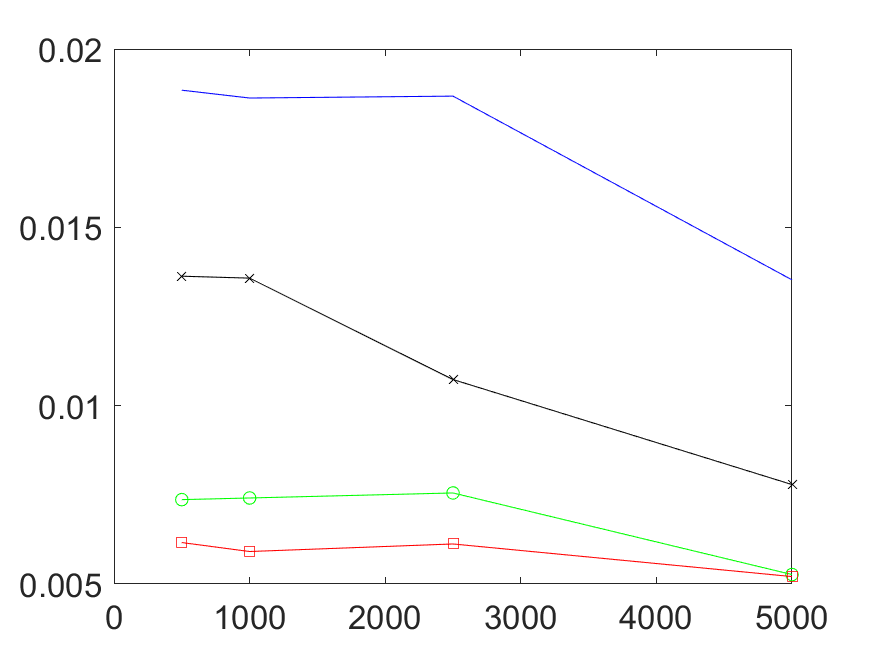}  \\
       & Scaled log-likelihood & MAE probabilities & average KS distance & MAE percent negative
\end{tabular}
\caption{Results for cases (1a) to (1c): {\tt GR} in blue, {\tt EM-GR} in green, {\tt EM} in black, {\tt BE} in red. } \label{fig:case1}
\end{figure}

In all cases the alternative specific constants are zero while the coefficient for the regressor vector is nonzero and distributed according to three different distributions. 
The results provided in Figure~\ref{fig:case1} show that in all three cases the adaptive grid algorithm {\tt GR} does not provide optimal results. This is to be expected due to 
the mis-specification of the error variance.
The likelihood values show that it fails to find solutions better than the true parameter on average in all cases (1a)-(1c), although the difference is small. 

Starting from this point also the algorithm {\tt EM-GR} does not provide good performance
-- while improving compared to {\tt GR} -- except in case (1b). {\tt EM} does a much better job. 
While {\tt EM} provides consistent estimates for the choice probabilities in all cases, {\tt EM-GR} only does so in case (1b) of normally distributed coefficient. 

The results also show that the error in the estimated distribution of $\beta$ does not appear to be decreasing in case (1a) for the {\tt GR} and the {\tt EM-GR} approach while it does so for the other methods.

\begin{figure}
\begin{tabular}{ccc}
\includegraphics[width=45mm]{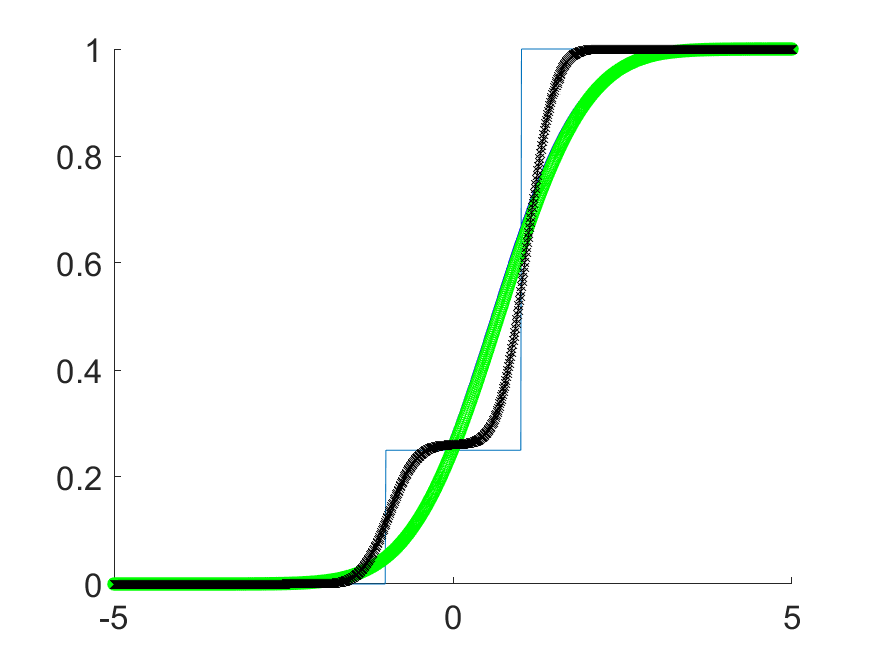} &   \includegraphics[width=45mm]{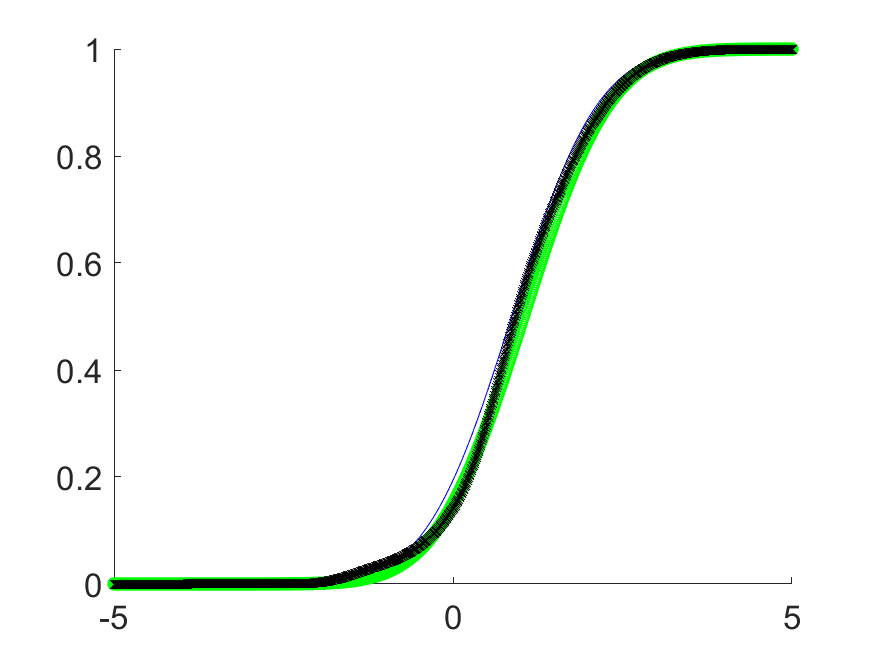} &  \includegraphics[width=45mm]{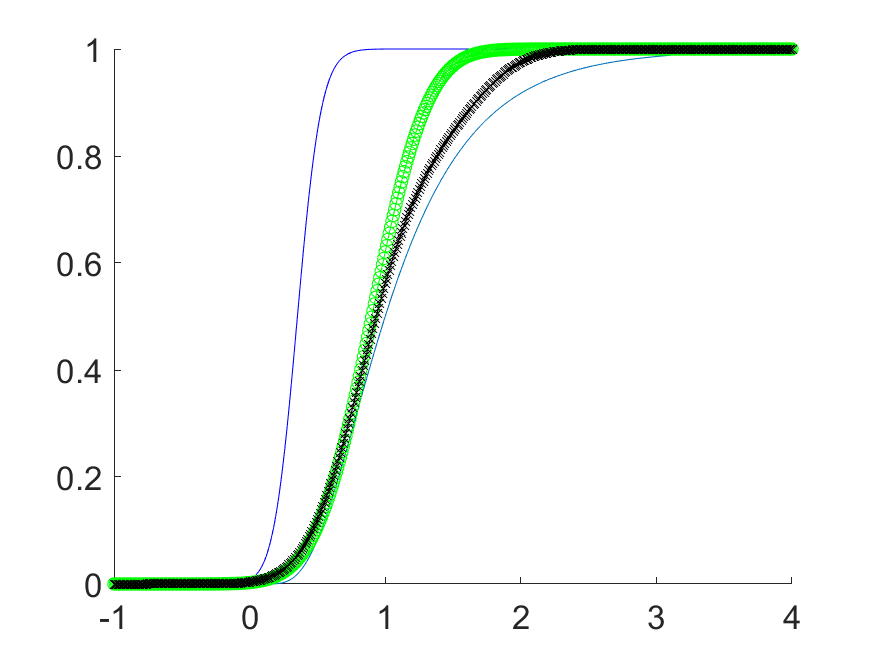}  \\
Case (1a) & Case (1b) & Case (1c)
\end{tabular}
\caption{Results of estimation for CDF for cases (1a) to (1c) for one simulation and sample size $T=5000$: {\tt GR} in blue, {\tt EM-GR} in green, {\tt EM} in black. } \label{fig:case1_cdf}
\end{figure}

With respect to the frequency of negative values of $\beta$ according to the estimated mixing
distribution note that in case (1a) {\tt GR} and {\tt EM-GR} do not appear to converge to the true
percentage. However, the difference is small (2.5\%). For case (1b) and {\tt GR} it is more
substantial with 4\%.  

As a final finding note that in case (1b) of a normal mixing distribution the method {\tt 
EM-GR} outperforms {\tt EM}. This is due to the fixed variance of the components in the 
second case which does not allow adaptation to the variance of the modelled random 
variable. 

\subsection{Results for Cases (2a)-(2b)}
Here the coefficient $\beta = 1$ is kept fixed and is not estimated -- fixing the scale of the utility -- while the two alternative specific constants are distributed randomly. 
This case can be viewed as alleviating the assumptions on the error term by allowing correlation as well as deviation from normality. In this case the mis-specification of the
error variance in the algorithm {\tt GR} is expected to be of minor importance as here 
the distribution of the ASCs can be used to include this feature in the model.

\begin{figure}
\begin{tabular}{c m{4.5cm} m{4.5cm}  m{4.5cm}}
Case (2a) &  \includegraphics[width=45mm]{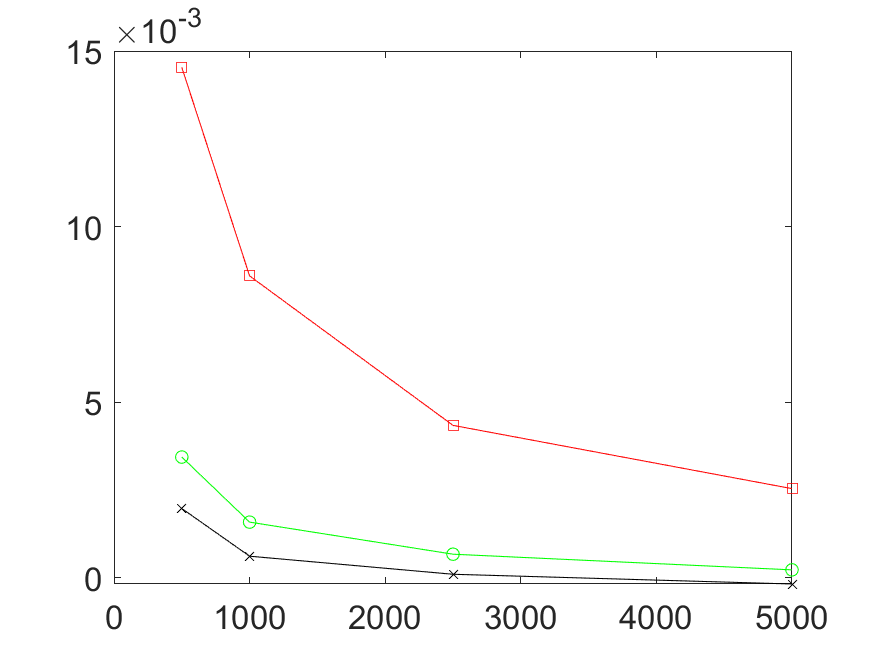} &   \includegraphics[width=45mm]{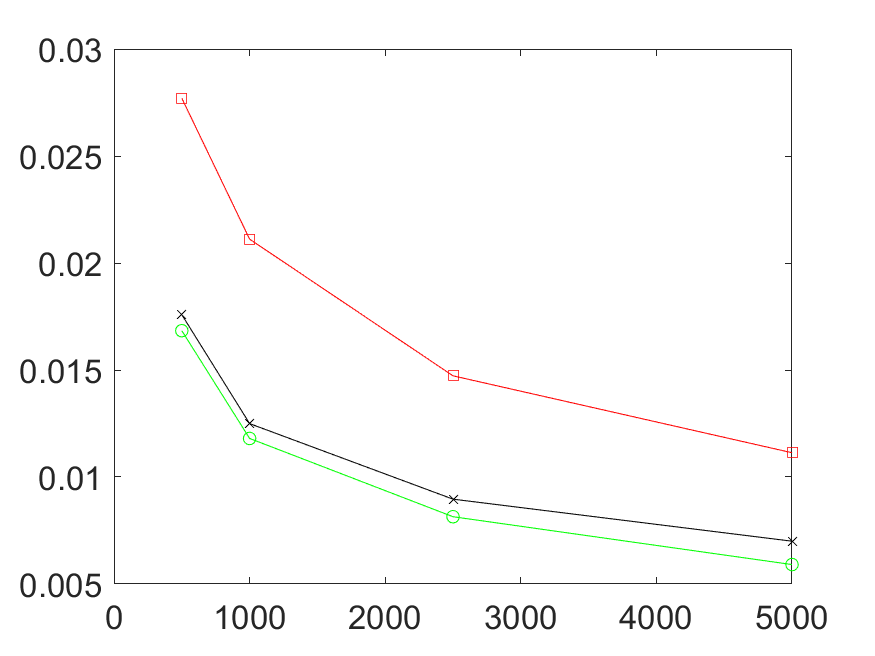} &   \includegraphics[width=45mm]{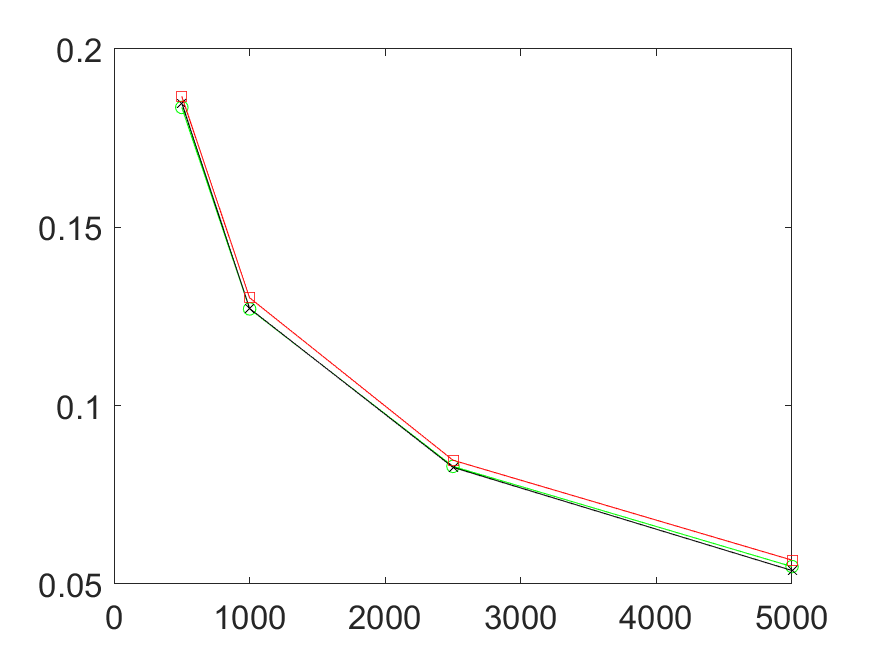}  \\
Case (2b) &  \includegraphics[width=45mm]{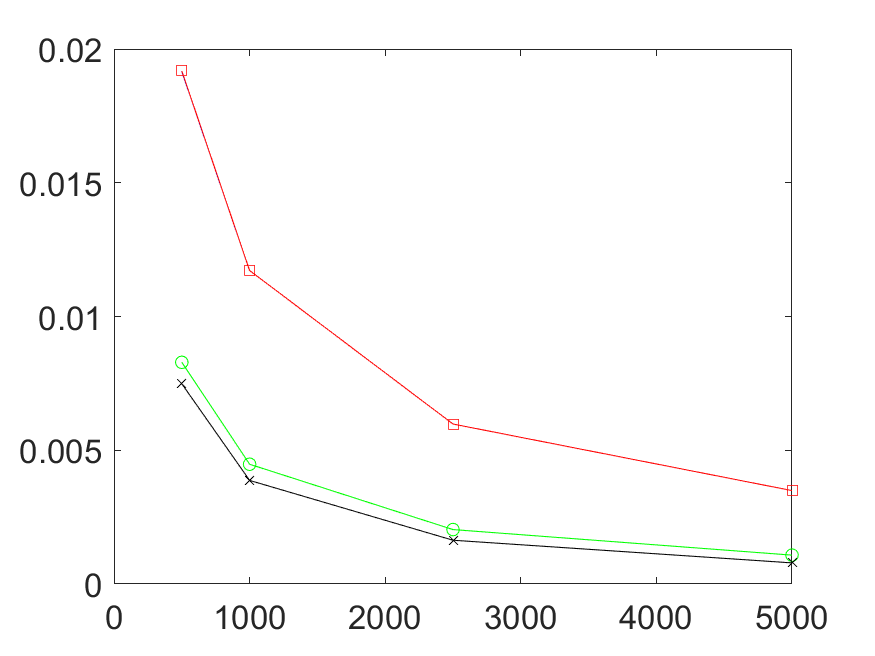} &   \includegraphics[width=45mm]{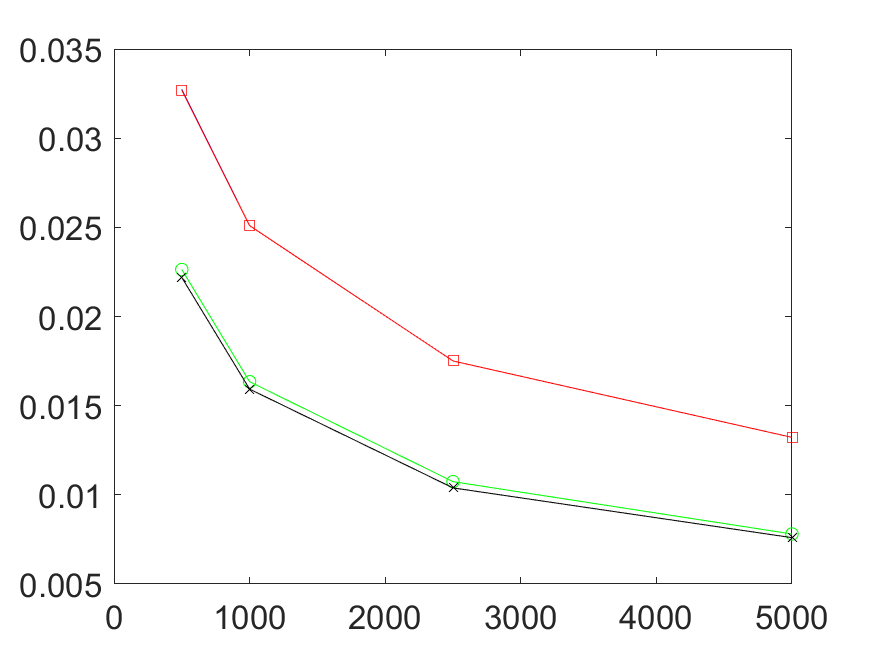} &   \includegraphics[width=45mm]{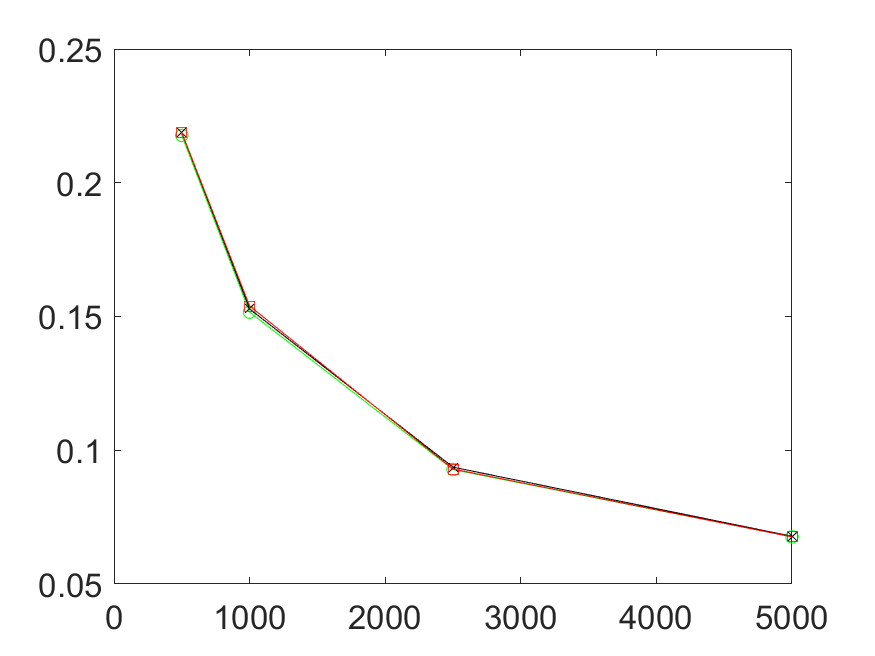}  \\
       & Scaled log-likelihood & MAE probabilities & average norm of difference in expectation
\end{tabular}
\caption{Results for cases (2a) and (2b): {\tt GR} in blue, {\tt EM-GR} in green, {\tt EM} in black, {\tt BE} in red. } \label{fig:case2}
\end{figure}

The results show that while the likelihood values are slightly larger for the {\tt GR}
method, the {\tt EM} based methods achieve a slightly better match of the choice probabilities. 
Note, however, that in these cases all estimates are very accurate with a mean average error of less than 3\% in all cases and less than 1\% in the largest sample size. 

In this respect it is interesting to see that taking the best fitting (in terms of
in-sample likelihood) algorithm results in higher in-sample likelihood but slightly worse
out-of-sample performance. 

Finally note that in these two cases (2a) and (2b) all algorithms achieve a high accuracy in the estimation of some quantities like the expected alternative specific constants value.
The third column of Figure~\ref{fig:case2} provides the average norm of the difference between the expectation of the estimated ASCs distributions and the true expectation. 
It can be seen that for all estimators as sample size increases the expectation is estimated with high precision. 
In this respect it is interesting to note that the expectation is estimated with the same level of accuracy while the estimated choice probabilities show some differences in accuracy.

%
%
%

\section{Conclusions} \label{sec:concl} 
In this paper we demonstrate that the NP-MLE approach reviewed in \cite{Lindsay} can be used in order to derive non-parametric estimators for the distribution of the mixed parameters in mixed MNL and mixed MNP models. 
Identification of the various models follows from the work of \cite{fox2012random}, where also estimation algorithms are found. 

We combine the NP-MLE literature with the corresponding estimation algorithms relying heavily on the structure of the NP-MLE problems with ideas from the theory of adaptive grids in order to alleviate the problems of the curse of dimensionality. 
The bottom line of the literature review is that the mixing distribution can be estimated consistently even when using latent class models with the number of classes tending to infinity
under suitable identifiability results involving sufficiently informative regressors. 

Introducing ideas from the density estimation we suggest to use a Gaussian mixture in order to represent the mixing distribution. 
Thus we obtain a number of estimation algorithms relying on NP-MLE using the EM algorithm to find the appropriate support points or relying on adaptive grids in order to involve new grid points. 

The simulations show that the adaptive grid algorithms work well in situations with a good approximation using the initial grid, that is situations close to normality. 
Otherwise the EM-based algorithms starting from a general grid work better in the investigated situations of low dimensionality.

Another interesting result -- which at least for us was not obvious from the outset -- is
the low sensitivity of the choice probability estimates as a function of the mixing
distribution. For cases (1a)-(1c) we obtained relatively good estimates for the choice probabilities even for biased model structures. 
This provides evidence that very large sample sizes are needed in order to obtain somewhat reliable information on the particular form of the mixing distribution.
However, some quantities like the expected coefficient can be estimated accurately even in moderate sample sizes. 

We also saw in the examples that the probability that a coordinate of $\beta$ takes on positive values can be estimated fairly accurate as well as the 
expectation of the mixing distribution. 
Further undocumented results also suggest that other quantities like the variance of the mixing distribution are harder to estimate: here very large samples
are needed in order to obtain meaningful estimates. 

During the simulations we witnessed already fairly large computation times for these small scale models. As an example for sample size $I=1000$ one simulation for case (2a) took 7.6 minutes on a standard laptop. 
Thus future work will be directed towards better choices for some of the parameters of the estimation algorithms in order to reduce the computational burden while retaining estimation
accuracy. 

\section*{Acknowledgements}

This work has been financed partly by the Deutsche Forschungsgemeinschaft (DFG, German Research Foundation) - Projektnummer 356500581 which is gratefully acknowledged. 

\bibliographystyle{plain}

\end{document}